\documentclass[aps,pre,reprint,groupedaddress]{revtex4-1}

\usepackage[pdftex]{graphicx}
\usepackage[english]{babel}
\usepackage[T1]{fontenc}
\usepackage{amsmath}

\usepackage[colorlinks,linkcolor=blue,citecolor=blue,urlcolor=blue]{hyperref}
\hypersetup{pdfauthor={P. A. Mu\~noz},
pdftitle={Kinetic turbulence in fast 3D collisionless guide-field magnetic reconnection}
}

\begin{document}

\title{Kinetic turbulence in fast 3D collisionless guide-field magnetic reconnection}
\author{P.~A. Mu\~noz}
\email{munozp@mps.mpg.de}
\affiliation{Max-Planck-Institut f\"ur Sonnensystemforschung, D-37077 G\"ottingen, Germany}
\altaffiliation{Also at Max-Planck/Princeton Center for Plasma Physics. \\ Now at
EW 215, Center for Astronomy and Astrophysics, Technical University Berlin, Hardenbergstra{\ss}e 36, 10623 Berlin, Germany}

\author{J. B\"uchner}
\affiliation{Max-Planck-Institut f\"ur Sonnensystemforschung, D-37077 G\"ottingen, Germany}
\altaffiliation{Also at Max-Planck/Princeton Center for Plasma Physics. \\ Now at
EW 215, Center for Astronomy and Astrophysics, Technical University Berlin, Hardenbergstra{\ss}e 36, 10623 Berlin, Germany}

\date{\today}

\begin{abstract}
	Although turbulence has been conjectured to be important for magnetic
	reconnection, still very little is known about its role in collisionless plasmas.
	Previous attempts to quantify the effect of turbulence on reconnection usually
	prescribed Alfv\'enic or other low-frequency fluctuations or investigated
	collisionless kinetic effects in just two-dimensional configurations and
	antiparallel magnetic fields.
	In view of this, we analyzed the kinetic turbulence self-generated
	by three-dimensional guide-field reconnection through force-free current sheets
	in frequency and wavenumber spaces, utilizing 3D particle-in-cell code numerical simulations.
	Our investigations reveal reconnection rates and kinetic turbulence with
	features similar to those obtained by current in-situ spacecraft observations of
	MMS as well as in the laboratory reconnection experiments MRX, VTF and
	\textsc{Vineta}-II.
	In particular we found that the kinetic turbulence developing in the course of
	3D guide-field reconnection exhibits a broadband power-law spectrum extending
	beyond the lower-hybrid frequency and up to the electron frequencies.
	In the frequency space the spectral index of the turbulence appeared to be close
	to -2.8 at the reconnection X-line. In the wavenumber space it also becomes
	-2.8 as soon as the normalized reconnection rate reaches 0.1.
	The broadband kinetic turbulence is mainly due to current-streaming
	and electron-flow-shear instabilities excited in the sufficiently
	thin current sheets of kinetic reconnection.
	The growth of the kinetic turbulence corresponds to high
	reconnection rates which exceed those of fast laminar, non-turbulent
	reconnection.
\end{abstract}

\maketitle

\section{Introduction\label{sec:intro}}
Magnetic reconnection is a fundamental process that converts magnetic energy into
kinetic energy and heat in laboratory, space and astrophysical
plasmas~\cite{Yamada2010,Treumann2013b}.
Though ubiquitous in the collisionless plasmas of the Universe, it is not clear, yet,
whether and which turbulence can enhance the energy conversion by reconnection.
In the past it has been conjectured that macroscopic (fluid) turbulence can enhance
the reconnection rate~\cite{Matthaeus1985,Lazarian1999}.
But the role of kinetic turbulence in collisionless magnetic reconnection
is much less explored.
Indeed, as is well known, current sheets (CSs), through which magnetic
reconnection takes place, contain a sufficient amount of free energy  which in
collisionless plasmas is released by instabilities at the smallest, kinetic scales.
In contrast to fluid (magnetohydrodynamics, MHD) turbulence, the universality and properties of
collisionless turbulence self-generated by magnetic reconnection, such as their scaling,
power law spectral index and spectral breaks, are not yet well understood.

Recent in-situ measurements often detected thin current sheets
formed in the turbulent space plasmas of the solar wind or Earth's magnetosheath, leading to
magnetic reconnection events generating heating and
dissipation~\cite{Retino2007,Sundkvist2007,Gosling2012,Perri2012,Osman2014,Wang2015s,Osman2015,Chasapis2015,Voros2016a,Yordanova2016}.
The small-scale turbulence in the
vicinity of those CSs can usually be associated to spectral breaks in the magnetic fluctuation
spectra near the
ion cyclotron frequency $\Omega_{ci}$. At larger scales (low frequencies),
there is the characteristic inertial range of the turbulent cascade, while above ion scales
 the turbulent spectra
shows a clear power law with spectral indices close to $-2.7$~\cite{Eastwood2009,Huang2012}.
However, the power-laws and spectral breaks near CSs are very similar to those measured in homogeneous turbulent solar wind plasmas~\cite{Alexandrova2009,Eastwood2009,Chen2013d}, and that is why it is not
well known how much reconnection contributes to the measured spectra.
In addition, the spectral breaks might not be universal and depend on several
parameters~\cite{Chen2014f,Bourouaine2012,Bruno2014,Boldyrev2015,Breuillard2016a}.
Note that similar spectra, but with different spectral indices, were also obtained
for other quantities such as the electric field~\cite{Bale2005,Chen2011d,Mozer2013,Matteini2016a},
density~\cite{Chen2012a,Chen2014,Safrankova2015} and higher order momenta of the distribution functions
such as bulk flow velocity and temperature~\cite{Podesta2006,Safrankova2013,Safrankova2016}.
On the other hand, laboratory experiments of magnetic reconnection
(MRX, VTF, \textsc{Vineta}-II)
obtained turbulent cascades as well but with different spectral indices and
spectral breaks near the lower-hybrid frequency $\Omega_{LH}=\omega_{pi}/\sqrt{1+\omega_{pe}^2/\Omega_{ce}^2}$ (where $\omega_{pe}$ ($\omega_{pi}$) is the electron (ion) plasma frequency and $\Omega_{ce}$ is the electron cyclotron frequency)~\cite{Carter2002,Fox2010,VonStechow2015a}.
These different spectral breaks indicate a change
in the physical nature of  turbulence,
depending only on the ions (in space plasmas) or with an influence of the electrons
(in laboratory plasmas).
In space, the first spectral break
observed in  steady and homogeneous solar wind turbulence
is usually explained by a turbulent cascade of
kinetic Alfv\'en waves (KAW)~\cite{Leamon1999,Passot2015}, whistler waves~\cite{Stawicki2001},
Landau damping~\cite{Leamon1999,Howes2008,Boldyrev2013}, ion cyclotron damping~\cite{Gary1999a}
or combinations of these mechanisms depending on parameters such
as the plasma-$\beta$~\cite{Boldyrev2015}.
There is even some evidence from measurements of homogeneous turbulence
 in the solar wind of a second spectral break near electrons scales and a
steeper power law spectral index beyond it~\cite{Sahraoui2009,Sahraoui2013a,Huang2014e}.
However, measurements at those high electron frequencies
are more difficult to obtain due to the instrumental limitations,
and other interpretations such as an exponential cutoff beyond electron scales
have also been proposed~\cite{Alexandrova2009}.
It is important to mention, however, that all those spectra measurements
in space plasmas are performed in the spacecraft frame of reference,
which gives a Doppler-shifted frequency $\omega_{sc}=\omega + \vec{k}\cdot \vec{v}_{sw}$,
where $\vec{v}_{sw}$ is the plasma flow (solar wind) speed.
In order to compare with theoretical predictions in the plasma frame of reference,
a transformation has to be carried out to the wavenumber domain $\vec{k}$ by
assuming the validity of Taylor's hypothesis~\cite{Taylor1938}
$\omega \ll \vec{k}\cdot \vec{v}_{sw}$. This implies a linear relationship
between the spacecraft frequency space
and the wavenumber domain in the plasma frame $k=\omega_{sc}/(|v_{sw}|\cos\theta)$, while the frequency spectrum
in the plasma frame $\omega$  remains unknown. In the last expression,
$\theta$ is the angle between the wavenumber $\vec{k}$ and solar wind velocity $\vec{v}_{sw}$, i.e.,
 $\vec{k}\cdot\vec{v}_{sw}=k|\vec{v}_{sw}|\cos\theta$.
This hypothesis is not valid for slow plasma flows or for high frequency dispersive
waves with a non-linear relation with the wavenumber~\cite{Klein2014,Klein2015b}.

Another important point related to terminology is that electron scales
in spacecraft measurements have a slightly different meaning than
in a pure frequency analysis in the plasma frame.
This is because it refers to the wavenumber $kd_e$ (electron skin depth) mapped to the frequency
space by using the previously discussed Taylor's hypothesis, resulting in
$\omega_{de}=v_{sw}/d_e$. This frequency is closer to the corresponding ion scales
$\omega_{di}=v_{sw}/d_i$ (mapping of the wavenumber related to the ion skin depth $kd_i$)
by a factor of the square root of the mass ratio $\sqrt{m_i/m_e}$
(also valid for the electron (ion) gyroradius $k\rho_e$ ($k\rho_i$))
compared to a direct frequency spectra, where
the ion and electron frequencies $\Omega_{ci}$ and $\Omega_{ce}$ are separated by $m_i/m_e$.

The properties of stationary and homogeneous kinetic turbulence
leading to localized magnetic reconnection were
numerically investigated using hybrid-PIC (Particle-in-Cell), gyrokinetic or Vlasov codes simulations~\cite{Howes2008b,Parashar2009,Greco2012,Valentini2014a,Franci2015a,Servidio2015,BanonNavarro2016b,Parashar2016a,Cerri2017,Franci2017a}.
These investigations showed that ion-scale CSs, where magnetic reconnection can take place,
 can form from decaying or driven turbulence, leading to heating, temperature
anisotropies and dissipation.
Fully-kinetic PIC code simulations of shear-driven or decaying turbulence
demonstrated that CSs also form at electron scales and contribute to the turbulent spectra~\cite{Wan2012,Wu2013e,Karimabadi2013b,Haynes2014,Wan2015,Groselj2018}.
A recent comparison of shear-driven turbulence with different physical and
numerical models also showed the formation of current sheets and hinted
towards the importance of reconnection in turbulence at different scales~\cite{Groselj2017}.
The idea that magnetic reconnection can have an important contribution to
turbulence has also been suggested by a number of recent theoretical
works in both collisional and collisionless
regimes~\cite{Howes2015b,Mallet2017b,Mallet2017a,Loureiro2017,Loureiro2017b,Boldyrev2017}
(see also the review \cite{Matthaeus2015}).
In spite of all those works assessing the role of multiple magnetic reconnection
events in turbulence, the opposite problem of the turbulence generated by
magnetic reconnection in a single isolated current sheet has been rarely characterized.
Although there has been some work analyzing the turbulence generated by
plasmoid magnetic reconnection within the MHD framework~\cite{Huang2016t},
the self-generated turbulence due to kinetic instabilities in collisionless
magnetic reconnection is much less known.
For example, \cite{Leonardis2013} is one of the few studies analyzing this problem using
3D fully-kinetic PIC-code simulations of reconnection, revealing the presence of
non-Gaussian statistics and multifractal structures associated with intermittency
and dissipation.
Note that all these simulations usually revealed the spatial turbulence spectra,
while the also relevant frequency spectrum have remained mostly unknown, which
is one of the purposes of this study.

On the other hand, the properties of instabilities in CSs
and their consequences for the kinetic turbulence generated during magnetic reconnection
were also investigated by using 3D PIC-code~\cite{Buchner1999} or
fully-kinetic 3D Vlasov-code simulations~\cite{Silin2003,Silin2003a}.
In particular, the role of  Buneman instability was
studied with Vlasov codes~\cite{Buchner2005a,Buchner2006}.
This instability is relevant to understand the consequences of the
self-generated turbulence in reconnection, because it can produce
anomalous resistivity and thus balance the reconnection electric field
associated to magnetic reconnection~\cite{Che2011,Che2017a}.
Even though in 2.5D fully-kinetic reconnection simulations this anomalous resistivity
could not be found~\cite{Munoz2017}, other 3D fully-kinetic reconnection
simulations provided some positive evidence~\cite{Drake2003,Che2011,Liu2013,Price2016,Price2017d}.

Although we do not attempt to make a direct comparison with spacecraft measurements
or check the validity of Taylor's hypothesis  under realistic conditions,
the need to study the properties of both the frequency and wavenumber
spectra at kinetic and dispersive scales generated by magnetic reconnection is clear.
In collisionless plasmas, the high-frequency kinetic turbulence is essential
for the balance of electric fields, and therefore for the
rate of magnetic reconnection, for energy dissipation
and heating~\cite{Buchner2007}.
\section{Simulations \label{sec:simulations}}
We investigated the kinetic turbulence,
self-generated in 3D collisionless reconnection,
and its consequences for the reconnection rate,
considering force-free equilibrium current sheets.
Those are closer to real and astrophysical CSs rather than
Harris-type CS, which require strong pressure gradients.
In force-free CSs with a guide field in the current  (our $z$-) direction,
the magnetic pressure is balanced by an electron shear flow in the $x$ direction,
while the thermal pressure is constant (see specific setup in~\cite{Munoz2015}).
We used the 3D fully-kinetic
PIC-code ACRONYM~\cite{Kilian2012}.
We illustrate our findings by presenting the results of a simulation run with an
ion to electron mass ratio $m_i/m_e=100$,
initially equal electron and ion temperatures ($T_i/T_e=1.0$),
a plasma beta $\beta_e=\beta_i = 2\mu_0 n_0 k_B T_i/B_T^2=0.016$,
a ratio of the electron thermal speed to the speed of light of $v_{th,e}/c=0.1$
and a relative guide field strength  $b_g=B_g/B_{\infty y}=2$,
where $B_{\infty y}$ is the asymptotic magnetic field (often abbreviated here as $B_0$).
The initial total magnetic field $B_T=B_{\infty y}\sqrt{1+b_g^2}$ is constant,
as well as the ion and electron number densities $n_0=n_i=n_e$.
Note that the plasma beta on the asymptotic magnetic field and guide fields
are, respectively, $\beta_{e,B\infty y}=\beta_{i,B\infty y}=0.08$ and
$\beta_{e,Bg}=\beta_{i,Bg}=0.02$.
Here, $v_{th,e}=\sqrt{k_B T_{e}}/m_{e}$ ($v_{th,i}=\sqrt{k_B T_{i}}/m_{i}$), and therefore the electron (ion) Larmor radius on the total magnetic field is
 $\rho_{e}=v_{th,e}/\Omega_{ce,BT}=(\sqrt{k_B T_{e}}/m_{e})/(eB_T/m_{e})$ ($\rho_{i}=v_{th,i}/\Omega_{ci,BT}=(\sqrt{k_B T_{i}}/m_{i})/(eB_T/m_{i})$).
This definition leads to a ratio of characteristic length scales $\rho_{i}/d_e=0.89$.
The current sheet halfwidth is $L=0.25d_i$.

In the case further discussed here the number of particles per cell (ppc) was 200 (100 per specie), with a total of $2.7\cdot 10^{10}$ particles in the simulation box.
The simulation box covers a domain
$L_x\times L_y\times L_z = 4\,d_i\times 8\,d_i\times 16\,d_i$,
where $d_i=c/\omega_{pi}$ is the ion inertial length ($\omega_{pi}$ is the ion plasma frequency).
The calculations were carried out on a mesh containing $256\times 512\times 1024$ grid points.
Periodic boundary conditions were chosen since we simulated two equivalent
but oppositely directed  current sheet flows.
For comparison with other studies of turbulence in the wavenumber domain, our
system allows the following minimum and maximum value of wavenumbers:
$k_{\parallel}d_i=[0.392-201]$ (or $k_{\parallel}\rho_{i}=[0.035-17.98]$), and $k_{\perp}d_i=[0.785-201]$ (or $k_{\perp}\rho_{i}=[0.07-17.98]$).
Here $k_{\parallel}=k_z$ (out-of-reconnection-plane direction, because of the
dominant guide field) and $k_{\perp}=k_y$ (along the reconnected component of the
magnetic field).

Reconnection is triggered by an initial magnetic field perturbation
with amplitude $0.07 B_{\infty y}$ for the corresponding  magnetic field components ($B_x$ and $B_y$).
This perturbation is narrowly localized in the current direction around $z=L_z/2$ and
with a long wavelength tearing mode in the $y$-direction, generating
a three-dimensional X-point.
\begin{figure}[!ht]
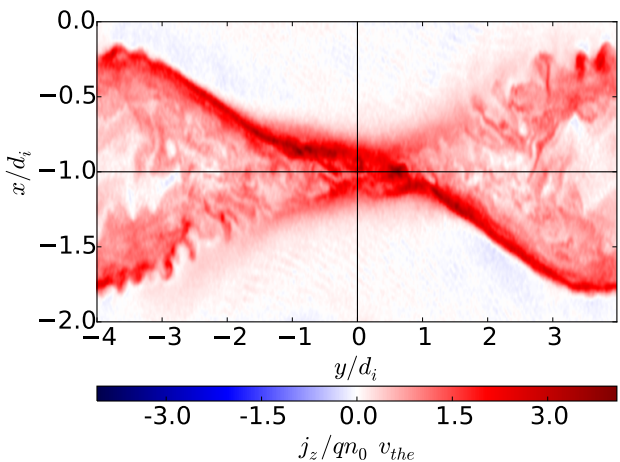

	\centering
		\includegraphics[width=0.99\linewidth]{{{fig1_new_contour}}}
		\caption{Spatial distribution of the current density $j_z$ at $t=13.5\,\Omega_{ci}^{-1}$ in the $x-y$ plane through $z=L_z/2$.
		\label{fig:contour}}
\end{figure}
\begin{figure}[!ht]
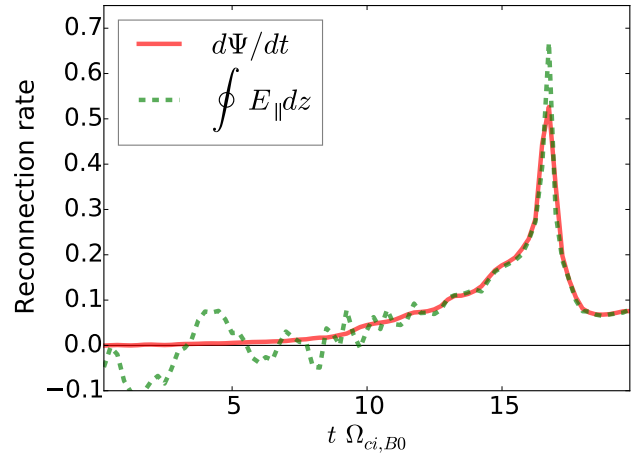

	\centering
	\includegraphics[width=0.99\linewidth]{{{fig2_new_rec_rate}}}
	\caption{Reconnection rate (normalized to $B_{\infty y}V_A/c$, with $V_A$ is the Alfv\'en speed on $B_{\infty y}$) calculated as the derivative of the reconnected flux between the X- and O-point of reconnection $d\psi/dt$ (red solid line), and as the parallel electric field integrated along the perimeter of
	the same region $\oint E_{\parallel} dz$ (green dashed line).
	\label{fig:rec_rates}}
\end{figure}
\section{Evolution of turbulence in reconnection \label{sec:turbulence}}
\begin{figure}[!ht]
	\centering
		\includegraphics[width=0.99\linewidth]{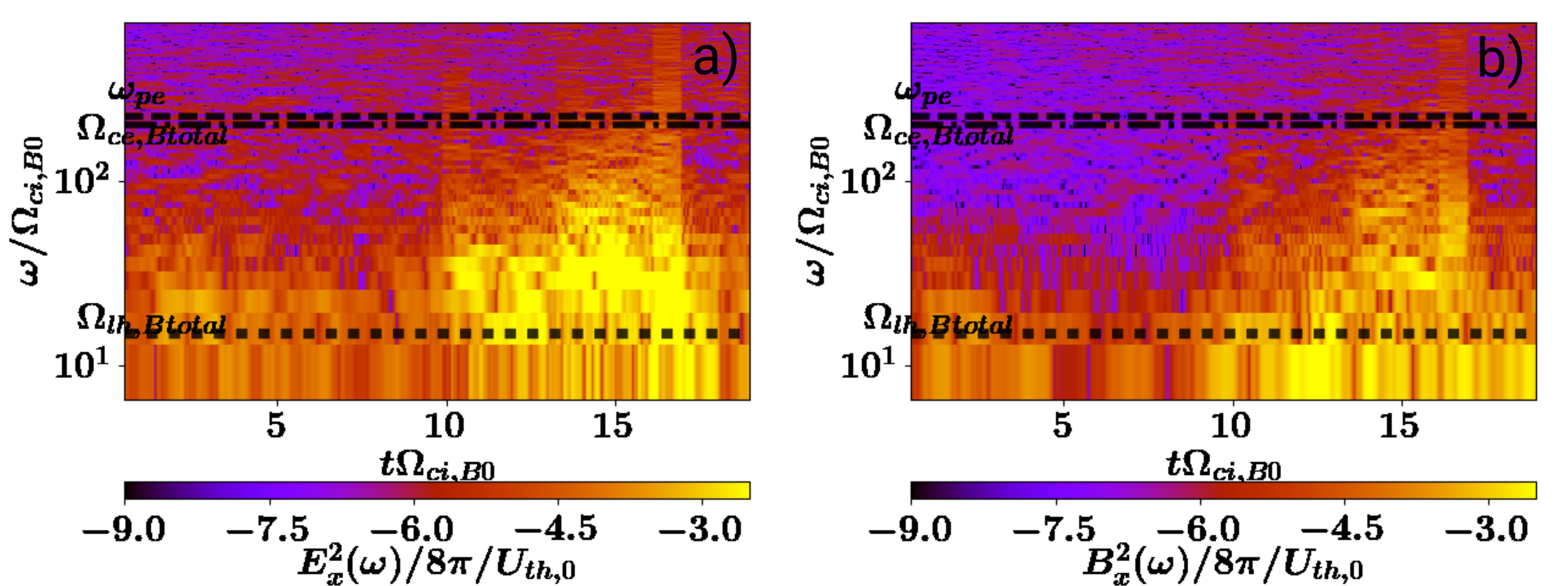}
	\caption{Temporal evolution of the spectra as a time-frequency representation (spectrogram) of the perpendicular electric $E_x^2$ (b) and magnetic  $B_x^2$ (b) fluctuations at the X-point of reconnection.
	Quantities are normalized to $U_{th,0}=(3/2)n_0k_BT_{e,0}$.
	The characteristic plasma frequencies are indicated by black horizontal dotted
	($\Omega_{LH}$), dashed-dotted ($\Omega_{ce}$) and dashed ($\omega_{pe}$) lines.
	\label{fig:spectrograms_xline}}
\end{figure}

Fig.~\ref{fig:contour} depicts the spatial distribution of the current density
distribution $j_z$ in the plane  $z=L_z/2$ at $t=13.5\,\Omega_{ci}^{-1}$,
after reconnection has fully  developed.
The Figure illustrates the well-known asymmetric structure of finite guide-field reconnection
and the spatial structure of turbulence.
Fig.~\ref{fig:rec_rates} shows the reconnection rate numerically calculated in
two different ways,
which, by definition, should be identical, both
characterizing the efficiency of reconnection.
$d\psi/dt$  (red solid line) is the rate of change of the total magnetic flux  calculated across a rectangle formed by  the X and O
lines of reconnection and the lines connecting $y=L_y/2$ and $y=0$.
This quantity should be equal to
$\oint E_{\parallel} dz$ (with $E_{\parallel}=\vec{E}\cdot\vec{B}/B$) represented with a green dashed line, the  parallel electric field integrated along the perimeter of the same rectangle.
The deviation between the two quantities is due to numerical errors caused by the PIC-code shot noise, which affects more the determination of the electric field rather than other quantities.
As Fig.~\ref{fig:rec_rates}) illustrates, reconnection starts to grow significantly only after $t\sim 10\,\Omega_{ci}^{-1}$,
reaches the limit of fast Petschek reconnection (0.1 in normalized units) at $t\sim13\,\Omega_{ci}^{-1}$, and further grows doubling that rate by  $t=15\,\Omega_{ci}^{-1}$.
But the peak at  $t=16.5\,\Omega_{ci}^{-1}$, with values of the reconnection rates
as high as $0.5 V_AB_{\infty y}$, is not due to only reconnection at the X-line,
but also due to the effects of the periodic boundary conditions:
the second current sheet starts to interact with the first current sheet
(the one studied here).
One effect of this is that the boundary of the magnetic island of the second current sheet is next to the X-line of the first one, limiting their growth and causing strong instabilities at the contact points due to the counterstreaming flows and possibly secondary reconnection sites.
A second effect is that the available magnetic flux incoming to each current sheet is drastically reduced, throttling the reconnection rates by a large amount.
In particular, the latter effect can be seen after $t\approx 16.5\Omega_{ci}^{-1}$
in Fig.~\ref{fig:rec_rates}, displaying a sharp decrease in the reconnection rates
to values below 0.1 $V_AB_{\infty y}$.
By $t\approx 18\Omega_{ci}^{-1}$ all the available magnetic flux is already exhausted and after $t\approx (20-21)\Omega_{ci}^{-1}$ reconnection stops.
Because of this, all the processes after $t\approx 15\Omega_{ci}^{-1}$ should already be affected by the direct interaction between the two current sheets and
are not representative of single X-line reconnection.
Note that the described evolution of reconnection in this system is dependent
on the simulation box size, especially along the current direction ($z$).
The reconnection onset and peak values of the reconnection rate are reached
later for longer boxes and the whole reconnection process is longer if
the system is long enough along the current direction.
The dynamic spectrum of the turbulence is depicted by Figs.~\ref{fig:spectrograms_xline},
which show the temporal evolution of the frequency spectrum of electric and magnetic fields
 in the direction perpendicular to the current flow direction at the
X-point of reconnection.
We obtained them by a short-time Fourier transform using a sliding Tukey window with an appropriate overlap and plotted as spectrograms.
Figs.~\ref{fig:spectrograms_xline} shows that until about  $t=10\,\Omega_{ci}^{-1}$,
significant turbulence is developed only below the lower-hybrid frequency $\Omega_{LH}$.
After that time both electric and magnetic
turbulence  strongly increase at kinetic scales up to the electron frequencies.
The turbulence broadening correlates well with the enhanced reconnection rates
(cf. Fig.~\ref{fig:rec_rates}).

\begin{figure}[!ht]
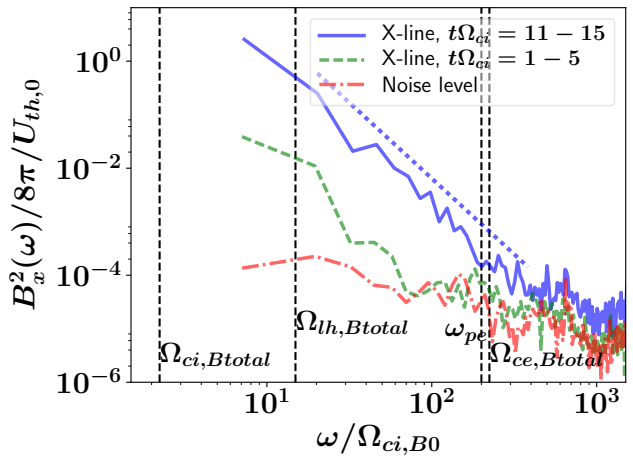

	\centering
		\includegraphics[width=0.99\linewidth]{{{fig4_freq_spectra_new}}}
	\caption{Spectrum of perpendicular magnetic field fluctuations $B_x^2$ at the reconnection X-point, for two time intervals. The raw simulation data is binned (see main text for details).
	The power spectra are normalized to $U_{th,0}=(3/2)n_0k_BT_{e,0}$.
	The numerical noise level is shown for comparison.
	The black dashed vertical lines indicate characteristic plasma frequencies.
	The diagonal dashed blue line represents the linear fit with spectral slope -2.8
	of the continuous blue curve, within the indicated frequency range.
\label{fig:time_spectra_xline}}
\end{figure}

\section{Frequency spectra \label{sec:frequency_spectra}}
In Fig.~\ref{fig:time_spectra_xline}, we show
the resulting frequency spectra of the perpendicular magnetic field fluctuations
$B_x$ at the X-line during two different characteristic time intervals:
$t\,\Omega_{ci} = 1-5$ and $t\,\Omega_{ci} =11-15$.
In order to diminish the noise level, we bin the raw simulation data and averaging every
eight data points (for both frequency and spectral power).
This is equivalent to an average over an interval of $14\,\Omega_{ci,B0}$.
Note that by binning the simulation data, the minimum resolved frequency becomes larger,
and closer to the lower-hybrid frequency $\Omega_{LH}$.
But it does not modify the spectral slope above  $\Omega_{LH}$,
which is our main interest here.

The comparison of these two spectra at the intervals
$t\,\Omega_{ci} = 1-5$ and $t\,\Omega_{ci} =11-15$
clearly demonstrates the development
of the high-frequency kinetic scale turbulence
above the lower-hybrid frequency
but also an enhanced spectral power close to $\Omega_{LH}$.
Note that Fig.~\ref{fig:time_spectra_xline} also shows the spectrum of the
(numerical) shot noise of the magnetic fluctuations (red dashed line)
 due to the finite number of particles used in the PIC-code simulations.
This noise spectrum was obtained at a location away from the CSs.
The plot indicates that fluctuations at
frequencies above $\omega_{pe}$ and $\Omega_{ce}$  (more specifically, above $\omega\gtrsim 400 \,\Omega_{ci,B0}$), shown as dashed vertical lines,
are due to numerical effects,
while the turbulence below  these electron frequencies
significantly exceeds the numerical noise level.
Above $\Omega_{LH}$, however, a clear steep power law spectrum $P=\omega^{\alpha}$
develops, with a spectral index $\alpha\approx-2.83$
which extends up to the electron frequencies $\omega_{pe}$ and $\Omega_{ce}$.
More precisely, we calculated the spectral index by means of
a least squares linear fit in the frequency range
$\omega/\Omega_{ci,B0}=17-400$ (for the interval $t\,\Omega_{ci} =11-15$), in order to consider all the frequency range above
 $\Omega_{LH}$ until the numerical noise level. This reference spectral slope
and its associated range is indicated as a dashed blue continuous line in Fig.~\ref{fig:time_spectra_xline}.
See Appendix~\ref{appendix} for a discussion about the effects
of the particle number and the choice of frequency range for the fitting on those results.

Note that in contrast to the usually used simpler spatial structure analysis of the
turbulence, we used a direct time-frequency-domain diagnostic of high cadence simulation data
by a stationary virtual probe located
at the X-point of reconnection, which provides the simplest and most general way
of analyzing the frequency turbulent spectra in this system.
This approach provides different insights in those kind of simulations,
while related work by \cite{Dmitruk2009,TenBarge2012} analyzed the frequency
spectra in homogeneous turbulence simulations.
Although this spectral index $-2.8$ is often measured by spacecrafts in
turbulent space plasmas undergoing reconnection between ion and
electron scales (roughly the frequency-mapped $kd_i$ to $kd_e$
by assuming Taylor's hypothesis)~\cite{Eastwood2009,Wang2015s},
a direct comparison is not appropriate,
because the spacecrafts are always in relative motion with respect
to the plasma frame. Instead, our method to obtain frequency spectra
actually compares better with laboratory experiments, where their probes
are stationary.
Furthermore, the compared range is not the same between simulations
and space observations:
the lower hybrid frequency is usually above the typical
frequency range accessible by space instruments, since
it approximately coincides with the location of the second spectral break
(at the frequency-mapped $kd_e$) if the frequency mapped $kd_i$ is close to $\Omega_{ci}$~\cite{Sahraoui2009,Sahraoui2013a,Huang2014e}.
Nevertheless, this range of frequencies is more easily accessible in
laboratory experiments,
which reveal a spectral break and a steep power law above $\Omega_{LH}$~\cite{Carter2002,Fox2010,VonStechow2015a}.

Note that it is important to obtain independently both frequency and wavenumber
spectra of fluctuations in order to get the plasmas dispersive properties
without any preliminary assumption such as Taylor's hypothesis,
which might not apply for high-frequency dispersive waves,
as in our simulations.
In view of this, we also calculated the spatial
spectra in our simulations. Thus, we can make a proper comparison with our resulting
frequency spectra, as well as to previous studies and observations or measurements.

\section{Wavenumber spectra \label{sec:wavenumber_spectra}}

	\begin{figure*}[!ht]
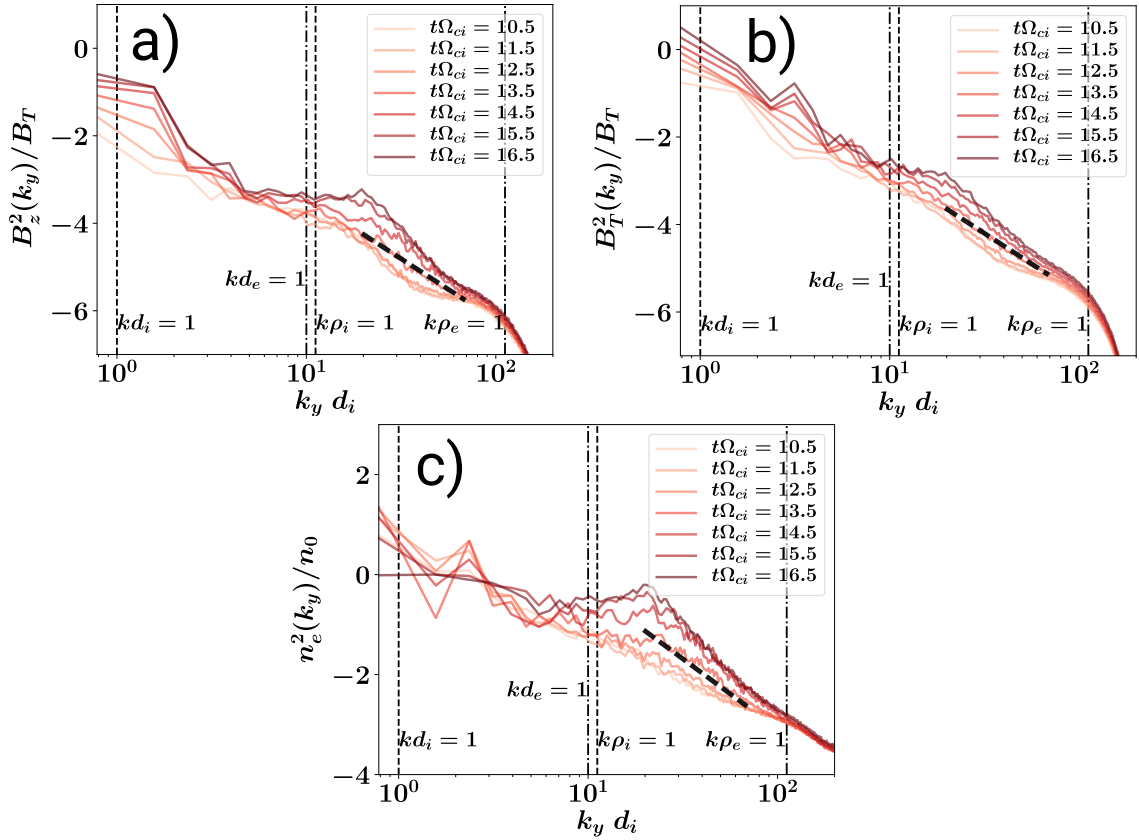

		\centering
			\includegraphics[width=0.85\linewidth]{{{fig5_comparison_spectra_ky}}}
			\caption{Spectral slopes of the $k_y$ power spectra for different times (in the range $t\Omega_{ci}=10.5-16.5$) and different physical quantities. a) $B_z^2$ magnetic field. b) total magnetic field $B_T^2$. c) $n_e^2$ electron density. The dashed oblique straight line indicates the reference slope $\alpha=-2.8$. \label{fig:comparison_spectra_ky}}
	\end{figure*}

The results for the calculation of the wavenumber spectra at
the center of the left CS $x=-1.0d_i$ and along the $y$ direction are shown in
Figs.~\ref{fig:comparison_spectra_ky}.
These figures show the power spectral density (PSD) of the
fluctuations in the parallel magnetic
field $B_z^2(k_y)$ (a) total magnetic field $B_T^2(k_y) = B_x^2(k_y) + B_y^2(k_y) + B_z^2(k_y)$ (b),  and electron density $n_e^2(k_y)$ (c).
Each wavenumber spectra is averaged along the out-of-plane direction $z$.
Common features for the magnetic field and density spectra are the
monotonously increasing spectral power as the times goes by,
a bump beyond $k=\rho_i^{-1}$ (more precisely, at $kd_i\sim 20$),
in particular for $\delta n_e$ and $\delta B_T$, and a
numerical steepening close to the grid scales for $k\rho_e > 1$.

Between $k\rho_i=1$ and $k\rho_e=1$, there are some ranges where a
straight line can be fitted. Therefore, we calculate spectral slopes $\alpha$
using a least squares linear fit of $P=k^{\alpha}$ for
all the available wavenumber data in a given range.
It is clear from Figs.~\ref{fig:comparison_spectra_ky}
 that those $k$-spectral indices are dynamical quantities depending on time,
loosely correlated with the value of the reconnection rate.
For reference, we indicate by a black dashed line the reference slope -2.8
close to the spectra obtained at $t\Omega_{ci}=13.5$.
The plots demonstrate a good fit
in the wavenumber range $k d_i= 20-80$.
For earlier times, the range for the linear fitting is moved to
smaller wavenumbers, $k d_i= 10-35$, because there is a clear flattening of the spectra
at about $k d_i \sim 40$, in particular of the $B_z$ component of the magnetic field fluctuations.
For late times ($t\Omega_{ci}\gtrsim13.5$),
we use the wavenumber range $k d_i= 20-80$ for the linear fitting, since this
includes the wavenumbers above the bump beyond $k\rho_i\ge1$, where the spectra
corresponds to a straight line.

\begin{figure}[!ht]
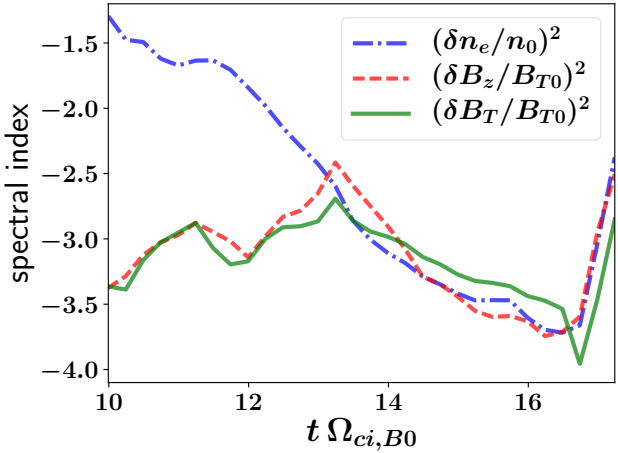

	\centering
		\includegraphics[width=0.99\linewidth]{{{fig6_spectral_slopes}}}
		\caption{Spectral slopes of the $k_y$ spectra for different times and different physical quantities.\label{fig:spectral_slopes}}
\end{figure}

The variation of the spectral indices with time is
summarized by the Fig.~\ref{fig:spectral_slopes}.
This Figure shows
that the slope of the electron density fluctuations continuously steepens
with time (see also Fig.~\ref{fig:comparison_spectra_ky}c),
reaching a maximum of $\alpha\approx -3.7$ at $t\Omega_{ci}\approx 16.5$.
At this time the reconnection rate is maximum
until it becomes determined by the interaction of the two current sheets
in the simulation domain.
Meanwhile, the parallel and total magnetic field fluctuations flatten
until  $t\Omega_{ci}\approx 13.2$.
Later they steepen again, with similar spectral indices,
and are also comparable in power to the electron density fluctuations.
Close to $t\Omega_{ci}\approx 13.2$, the spectral
indices reach the range  $\alpha=[-2.8, -2.5]$.
At this same time,
the normalized reconnection rate becomes  $0.1$
(c.f. Fig.~\ref{fig:rec_rates}).
The spectral indices become $\alpha=[-3.5, -3.3]$ at $t\Omega_{ci}\approx 15$,
where the energy conversion rate is $\sim 0.2$  (normalized reconnection rate),
before reconnection becomes affected by the second current sheet in the simulation domain
at $t\Omega_{ci}\approx 16.5$.

Therefore, the varying value of these spectral index slopes in the wavenumber domain
probably indicates that the kind of turbulence  developed due to non-steady reconnection,
dominated by instabilities, also changes during the course of reconnection.

We also analyzed the wavenumber spectra of
density and magnetic fluctuations along the out-of-plane direction $z$
(mostly aligned with the dominant guide field) in the region near the X-line.
A similar analysis
 also reveals a power law spectra
with similar variable spectral indices (and in approximately the same range).
Those wavenumber spectra, however, do not show a clear spectral break near
$k\rho_i=1$ and they also display a shorter turbulent cascade when no spatial average is used
(plots not shown here), since the noise level is higher.
Note that if an average of the wavenumber spectra in $k_z$ along the current sheet ($y$ direction)
would be performed, it would make to distinguish and fit a power law more difficult,
since the inhomogeneity and general features
of turbulence in the outflow region
are very different from those near the X-line, averaging out
different kind of processes.
In contrast, an average of the power spectral in $k_y$ along $z$ (the one used here)
is more consistent because it is similar among different slices.

The spatial spectral indices of the self-generated magnetic turbulence
along the CS are similar to those measured in space plasmas close to $\alpha=-2.7$, but only
when the (normalized) reconnection rates are close to 0.1.
Those spectral indices in the wavenumber domain are also within the same range
as the frequency spectral slope measured here for a stationary
probe at the X-point (see Fig.~\ref{fig:time_spectra_xline}).
However, this similar slope in $\omega$ and $k_y$ does not
necessarily indicate the presence of non-dispersive waves with
a linear dependence between $\omega$ and $k_y$,
since we verified
that the $\omega$-spectrum is different in the outflow regions
of reconnection: far from the X-line, there are less
turbulence and, therefore, the magnetic frequency spectrum does not
develop a spectral index as steep as in the X-line.
Since the $k_y$ spectrum considers equally all these regions
with different properties in the $\omega$ domain,
a dispersion relation $\omega$ - $k_y$ cannot be inferred uniquely
from a single sampling point in time.
Moreover, based on different 2D simulations for a similar
parameter range (with a higher output cadence),
dispersion relations $\omega$ - $k_y$ hint to the presence of dispersive
waves in the whistler branch with a quadratic dependence $\omega\propto k_y^2$.
Nevertheless, more work is needed to clarify if
the similarity of $\omega$ and $k_y$ spectral indices
is the result of our parameter range or a more generic
characteristic of this kind of turbulent system.

One of the goals of other works analyzing
the turbulence at kinetic scales is identifying whether the turbulent
fluctuations can be classified as due to  KAWs or
whistler waves~\cite{Chen2013d,Boldyrev2013,Cerri2016}.
The identification criteria is based on asymptotic formulas
leading to dispersion relations and associated transport ratios related to the
compressibility of fluctuations.
However, those expressions require $\omega\ll k_{\perp} v_{th,i}$ for KAWs
and $k_{\perp} v_{th,i} \ll \omega\ll k_{z} v_{th,e}$ for whistler waves
(see Fig.~1 of Ref.~\onlinecite{Boldyrev2013}), which are not well
satisfied in our case. One of the most important reasons is that
many of those formulas are derived for conditions $\beta\sim 1$, which do not apply
well in our simulations with small plasma-$\beta$, in addition to the use
of an artificially small mass ratio and simulation domain sizes.

\section{Instabilities leading to turbulence \label{sec:instabilities}}
\begin{figure}[!ht]
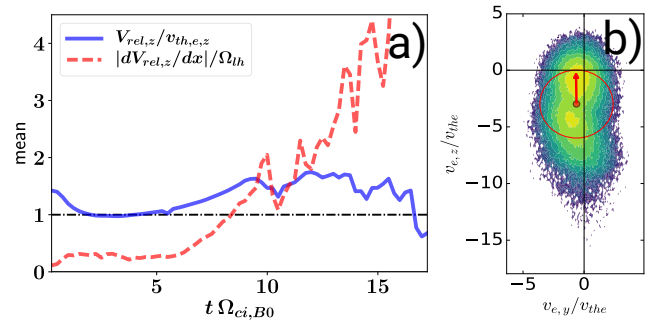

	\centering
	\includegraphics[width=0.99\linewidth]{{{fig7_instabilities_new}}}
	\caption{a) Time histories of quantities associated to streaming $V_{rel,z}$ (solid line) and shear flow $|dV_{rel,z}/dx|$ (dashed line) instabilities. The diagnostic corresponds to their mean value along the X-line of reconnection (average along $z$-direction).  The horizontal dashed line depicts a lower order estimation of the threshold of the Buneman instability.
	b) Electron distribution function in the plane $v_y-v_z$, taken at the X-line and near $L_z/2$, in a cubic region of size $0.1\,d_i\times 0.3\,d_i\times  1.0\,d_i$. The red arrow indicates the direction of the local magnetic field centered at the local bulk drift speed (red point). The red circle has a radius of $3v_{th,e}$, indicating approximately the initial distribution (99.7\% of the particles are inside of three standard deviations for a drifting Maxwellian with thermal spread equal to $v_{th,e}$.
	\label{fig:instabilities}}
\end{figure}

The broadband turbulence at the X-line self-generated by reconnection,
enhancing the spectral power between the lower hybrid and electron frequencies,
is caused mainly by a (streaming) Buneman instability~\cite{Buneman1958}.
The source of free energy of this instability is the relative
streaming of the current-carrying electrons with respect to the ions.
We verified this conjecture by investigating
the evolution of
the drift speed $V_{rel,z}$ along the X-line of reconnection,
with $V_{rel,z}=V_{i,z}-V_{e,z}$, where $V_{i,z}$ ($V_{e,z}$) is the ion (electron) drift speed along $z$.
As one can see in Fig.~\ref{fig:instabilities}a), initially,
in the thin CS $V_{rel,z}$ already  slightly exceeds
the initial electron thermal speed $v_{th,e}$.
This causes an initial (parallel) plasma heating, i.e.,
the electron thermal speed $v_{th,e,z}$ increases.
The marginal Buneman instability criterion $V_{rel,z}\sim v_{th,e,z}$,
however, is not reached again
until $t\sim 9\,\Omega_{ci}^{-1}$.
This way $V_{rel,z}/v_{th,e,z}$ can again increase above the threshold
of the Buneman instability.
Exactly at that time the broadband kinetic turbulence starts to develop.
After that the electron heating continues
while the electron-ion drift speed now grows even faster than
the electron thermal velocity.
This keeps the CS Buneman-unstable until boundary effects starts to play
a significant role close to $t \Omega_{ci} \sim 16.5$.
Note that previous 3D magnetic reconnection studies reported similar
Buneman-type instabilities and the generation of current filaments
in the current density along the $z$ direction~\cite{Drake2003,Che2011}.
The Buneman streaming instability is not effective in
2.5D magnetic reconnection in which, therefore,
no high-frequency turbulence near $\Omega_{LH}$ develops~\cite{Munoz2016a}.

As a second contribution to the broadband kinetic turbulence,
 which also enhances reconnection,
an electron shear flow instability is excited at the reconnection site.
Fig.~\ref{fig:instabilities}a) shows the gradient of the
current-aligned electron flow across the CS,
$dV_{e,z}/dx$.
It strongly grows after $t\Omega_{ci} \gtrsim 13$ when
it exceeds the threshold of the electron-ion hybrid (EIH)
instability  $dV_{e,z}/dx\gg \Omega_{LH}$~\cite{Ganguli1988,Romero1992}.
The EIH instability is a kinetic branch of the Kelvin-Helmholtz instability
which enhances the plasma turbulence near the lower-hybrid frequency.
The kinetic shear flow instability criterion is fulfilled, however,
only after the Buneman streaming instability has already developed.

\begin{figure}[!ht]
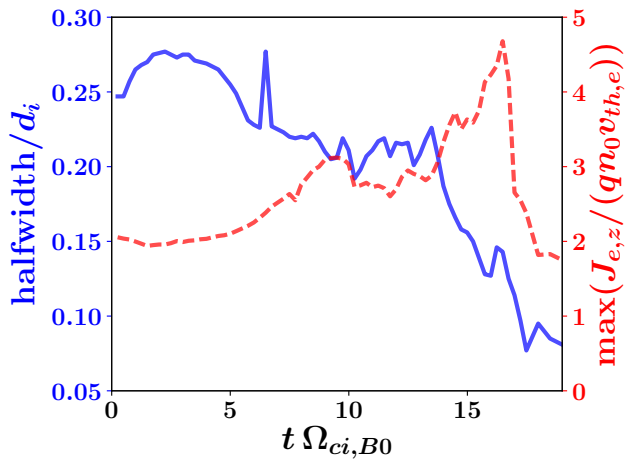

	\centering
		\includegraphics[width=0.99\linewidth]{{{fig8_cs_thinning}}}
		\caption{Time history of two characteristic quantities
		related to the current sheet evolution, calculated
		at the center $y=0$, $x=-1d_i$.
		Blue continuous line: current sheet halfwidth (values in the left axis). Red dashed line: maximum of $J_{e,z}$ (values in the right axis).
		See further details in the text.
	\label{fig:cs_thinning}}
\end{figure}

The most active period of both instabilities ($t\Omega_{ci}\gtrsim13.5$),
described above, it is also correlated with a fast thinning of the current sheet.
Fig.~\ref{fig:cs_thinning} shows the evolution of the halfwidth and
maximum of the electron current density $J_{e,z}$.
We choose to diagnose this quantity and not $J_{z}$ because most
of the current is carried by the electrons, both initially and also
later during the course of the CS evolution.
An ion current sheet also
forms self-consistently, but its contribution to the total current
is much smaller and it is also much broader.
We calculate the quantities shown in Fig.~\ref{fig:cs_thinning}
in the $x-y$ reconnection-plane along a cut in the
$x$ direction through the center of the first current
sheet ($y=0$). This is approximately the location of the X-point.
We also average $J_{e,z}(x,y)$ along the out-of-plane direction $z$.
Thus, the maximum value of the average $\bar{J}_{e,z}(x,y)$ is
obtained from calculations along this cut as shown in Fig.~\ref{fig:cs_thinning} (the red line).
On the other hand, we fit the function $f(x)=C + A\cosh^{-2}(x/\lambda)$
($A$, $C$ and $\lambda$ are all fitting variables) to the same $x$-cut in order
to get the halfwidth $\lambda$ of the electron current sheet.
This quantity is shown in Fig.~\ref{fig:cs_thinning} by a blue line.
Note that due to the averaging along the $z-$direction, the actual halfwidth
of the current sheet at given $z$-slices can be smaller or larger than that value.

This way, Fig.~\ref{fig:cs_thinning} shows that the current sheet halfwidth
quickly readjusts due to the initial perturbation and
the lack of exact kinetic equilibrium of
this force-free current sheet.
Also, the initial current sheet is slightly Buneman unstable
(c.f. Fig.~\ref{fig:instabilities}).
Later, when reconnection is laminar
($t\Omega_{ci}\lesssim 13.5$),
the current sheet halfwidth does not change much away from $0.2d_i$.
But during the non-linear evolution ($t\Omega_{ci}\gtrsim 13.5)$, when filaments appear, the reconnection rate is greatly enhanced and the spectral indices
of the magnetic fluctuations grow beyond $-3.0$.
Then, the halfwidth of the current sheet quickly
decreases until it reaches values as low as $0.08d_i$.
Meanwhile, the maximum $J_{e,z}$ grows steadily from the beginning,
stays more or less constant during the period of laminar reconnection,
to quickly grow, finally, during the nonlinear stage of reconnection.

Those instabilities and their turbulence also contributes
to the generation of significant deviations
in the electron velocity distribution function (EVDF)
from the initial drifting Maxwellian.
Beams are developed, temperature anisotropies and even a non-gyrotropy of the electron pressure tensor.
This is illustrated in Fig.~\ref{fig:instabilities}b),
showing a cut through the EVDF in the plane $v_y-v_z$ obtained near the X-line of reconnection.
A double peak is clearly visible in the EVDF,
 separated by about $4v_{th,e}$.
It is due to the interaction of two counter-streaming beams,
providing free energy for a two-streaming instability as well.
The origin of those beams can be understood as follows.
Originally, the electron distribution function is quite close to a shifted
Maxwellian with a drift speed (along $z$) close to the electron drift speed,
which is part of the initial conditions sustaining the current sheet.
The reconnection dynamics pulls magnetic flux and
thermal electrons (zero drift speed along $z$) from the upstream region into the X-line,
bringing together those two populations with a relative drift speed
between them. But reconnection also generates a reconnection
electric field which accelerates electrons, forming a beam-like
population with a very high drift speed and a non-thermal population as well.
This also produces an elongated plateau in the $-v_z$ direction,
which is a consequence of the non-linear evolution of the Buneman instability,
leading also to beam-driven lower-hybrid instabilities~\cite{McMillan2006,McMillan2007},
contributing to the turbulence near $\Omega_{LH}$.

Thus, while the beam-driven lower-hybrid instability is responsible mainly for
the turbulence near the lower-hybrid frequency,  Buneman and two-streaming instabilities
are behind the high frequency kinetic turbulence.
Since those instabilities should act simultaneously, it is not straightforward to
disentangle their individual effects, considering also they should be
mainly observed in their saturated state because of their large growth rate
and their source of free energy continuously being supplied by magnetic reconnection.
This high-frequency turbulence  might also quickly change the shape of the
EVDFs, but a detailed discussion about those effects is outside of the
scope of this paper.

\section{Kinetic turbulence and reconnection rates \label{sec:turbulence_reconnection_rates}}
In our simulations, the presence of high-frequency kinetic turbulence
is correlated with enhanced reconnection rates reaching
$0.2V_AB_{\infty y}$ and up to $0.5 V_AB_{\infty y}$.
Although we did not prove a causal relationship, there is some
evidence supporting that this association is not only a coincidence.
First of all, as has been established for a long time, normalized
reconnection rates close to 0.1 are typical for fast reconnection
in Harris or force-free current sheets and
within a wide range of parameters and physical models, more or less
independent on the dissipation mechanism
(see, e.g., Refs.~\cite{Comisso2016c,Liu2017w,Cassak2017c} and references therein).
We found here, though, that the reconnection rate can be significantly enhanced by Buneman turbulence, similar to the findings of~\cite{Che2017a}.
This can be interpreted as Buneman turbulence caused anomalous resistivity
balancing the reconnection electric field in the framework
of a generalized Ohm's law.
This requires relatively thin current sheets and fully 3D considerations.
That is why such an enhanced reconnection rate
was not commonly found in previous simulation studies, but
it is within the parameter regime of our study.
Therefore, the fact that Buneman instability is present in our simulations and reconnection rates are well above 0.1
hint towards the relation
between reconnection rates and this kind of Buneman turbulence.

\begin{figure*}[!ht]
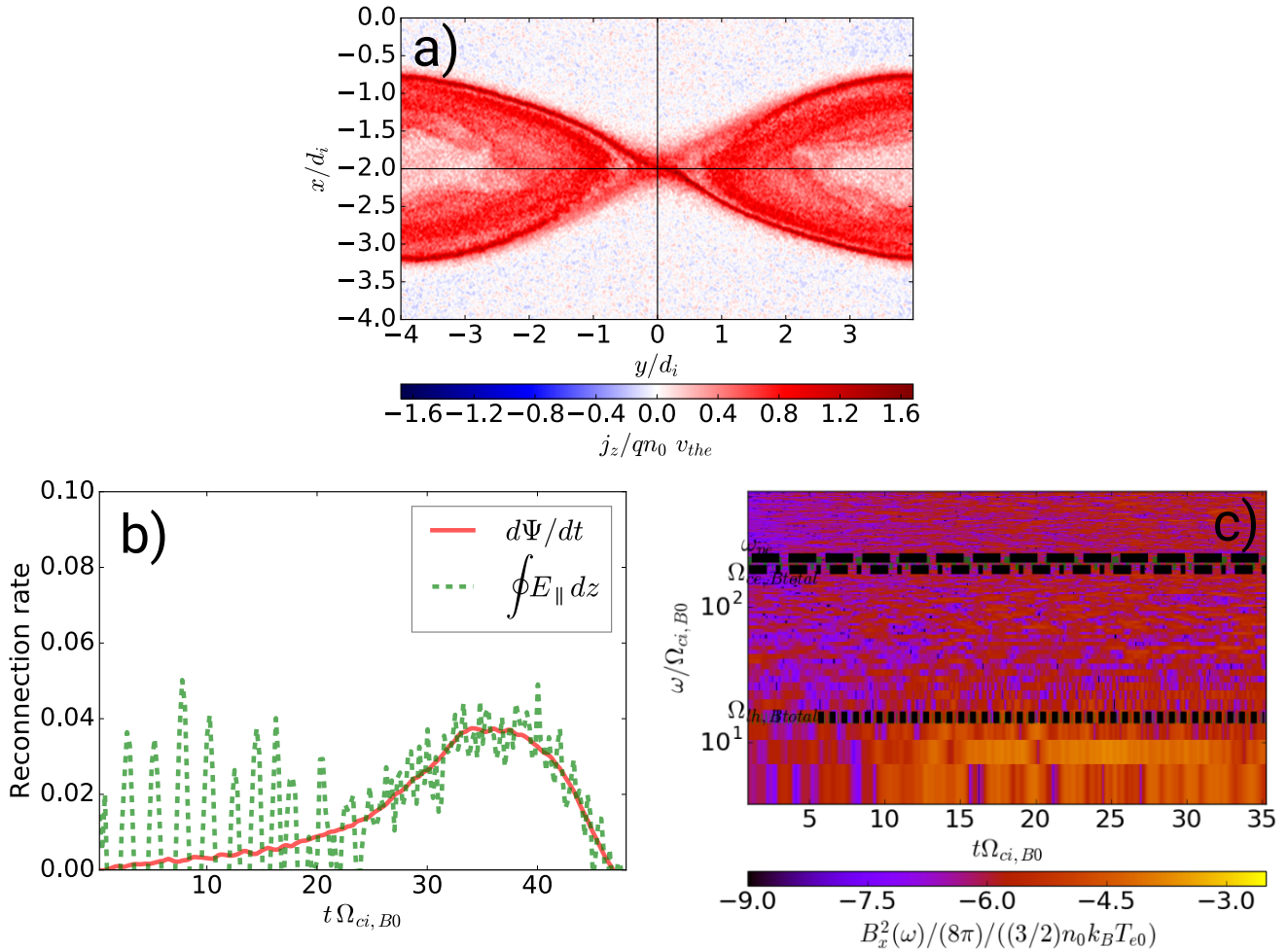
\centering
		\includegraphics[width=0.99\linewidth]{{{fig9_double_box}}}
	\caption{Results for a simulation run with a current sheet
	twice as thick and twice as large across the current sheet
	compared to the original run.
a) Current density $J_z$ at $t\Omega_{ci}=40$
b) Normalized reconnection rates.
c) Spectrogram of magnetic fluctuations.
\label{fig:double_run}}
\end{figure*}

To support the relation between self-generated
(mainly Buneman) turbulence and reconnection rate, we also
simulated thicker current sheets (e.g., exactly the double: $L=0.5d_i$)
for otherwise identical parameters.
In order to keep the separation constant between the current sheets,
we also increased the simulation box length across $x$ to twice its
original value.
We should mention that by changing the current sheet thickness
we are also modifying the stability properties of this system,
and so both simulation runs are not completely equivalent.
It was found previously that the
linear growth rates of the collisionless tearing mode are strongly reduced for thicker
current sheets~\cite{Pritchett1991,Brittnacher1995}.
In a just two times thicker current sheet
the reduced magnetic field shear implies a
relative electron-ion streaming speed below the threshold
of the Buneman instability ($v_{th,e}$). As a consequence, no Buneman
instability is triggered and the X-line does not become turbulent.
This can be seen in Fig.~\ref{fig:double_run}a) displaying
the out-of-plane current density $j_z$ at fully developed reconnection.
Note that the onset of reconnection is delayed because of the thicker current sheet and the larger simulation box size.
Fig.~\ref{fig:double_run}b) shows that reconnection rates are strongly reduced,
with their peak value close to $0.05 B_{\infty y}V_A$.
This reduction of the reconnection rate is associated with weaker
magnetic fluctuations, as it can be seen in Fig.~\ref{fig:double_run}c)
showing the spectrogram of the perpendicular magnetic field.
Different from the original run (see Fig.~\ref{fig:spectrograms_xline}),
it is clear that there is no development of a turbulent cascade and the spectral
power is not enhanced at all above the lower hybrid frequency $\Omega_{LH}$.
This supports the conjecture that
Buneman turbulence enhances reconnection.

\section{Conclusions \label{sec:Conclusions}}
We have analyzed the properties of self-generated kinetic turbulence
by 3D guide field reconnection
in both, frequency and wavenumber, domains.
In the course of reconnection, the self-generated turbulence
starts to grow near the lower-hybrid frequency.
Later, a broadband spectrum above the lower-hybrid frequency
and up to electron frequencies forms, exhibiting
a power law with a spectral index of $\alpha\sim-2.8$.
Different from previous investigations,
we obtained this power-law spectrum of perpendicular magnetic fluctuations directly
in the frequency domain for a stationary probe at the X-point
of reconnection.
For comparison purposes, we also analyzed the wavenumber spectra
in the direction perpendicular to the magnetic field.
This also reveals a power law spectra with a very similar spectral
index of $\alpha\sim-2.8$ for $k>\rho_i^{-1}$, but only
at times when the normalized reconnection rates are close to $0.1$.
This wavenumber spectral slope further steepens correlated with
enhanced reconnection rates above that value.
It is also associated with a
spectral break close to $k>\rho_i^{-1}$.
The similar slope of the $\omega$ and $k_y$ spectra does not
necessarily indicate the presence of non-dispersive waves with
a linear dependence between $\omega$ and $k_y$:
we verified that the $\omega$-spectrum is different in the outflow regions
of reconnection.
Those results cannot be directly compared with space
observations, but some of the characteristics of this kinetic turbulence
have some points in common, as well as with laboratory experiments.

The turbulence near the lower-hybrid frequency
is due to kinetic instabilities
driven by the streaming of the current carriers,
the electrons, their beams, and shear flows.
First the Buneman streaming instability starts
and later a kinetic electron shear flow instability takes over.
These two unstable modes might become coupled~\cite{Che2011b}.
The period of maximum activity of those unstable waves
correlates with a fast thinning of the current sheet.
The sources of free energy (electron currents and shear flows)
are typical for guide field reconnection.
This is in contrast to the limiting case of antiparallel reconnection,
where anisotropy-driven and pressure-gradient-driven instabilities prevail.
Our results indicate that simulations of magnetic reconnection need to be three-dimensional
to accurately describe the intrinsic 3D self-generated turbulence in a real physical current sheet:
2D setups cannot reproduce all the fluctuations and unstable waves seen in realistic environments.

We also provided some evidence that the high-frequency kinetic turbulence
generated by streaming and shear flow instabilities is correlated
with enhanced reconnection rates.
Usually, $0.1 V_AB_{\infty y}$ is considered to the
be the rate of fast reconnection.
However, we showed here that the rate of reconnection
through collisionless thin current sheets can be enhanced up to
$(0.2-0.5)V_AB_{\infty y}$ in the presence
of Buneman instability.
This was also found in a different study~\cite{Che2017a}.

By means of an additional simulation with a thicker current sheet,
where Buneman instability is not excited, we showed that the
consequent lack of high-frequency turbulence is correlated with weaker reconnection rates
on the order of $0.05 V_AB_{\infty y}$.
A more concrete and causal proof of this statement would
exceed the scope of this paper.

For larger ion to electron mass ratios and initially thicker current sheets,
the properties of the dominant instabilities might change.
It is very likely that a broadband kinetic turbulence
will nevertheless be excited and affect the reconnection process,
as laboratory experiments and in-situ observations have shown.
Starting with the current space mission MMS~\cite{Burch2016}
as well as by upcoming new laboratory experiments like FLARE at Princeton,
also higher (electron) frequency turbulence will become observable
which might compare better with our simulation results.

\begin{acknowledgments}
	We acknowledge the \textit{Verein zur F\"orderung kinetischer Plasmasimulationen e.V.}
	for developing the ACRONYM code and especially Patrick Kilian for his helpful discussions and valuable comments.
	We further acknowledge the Max-Planck-Princeton Center for Plasma Physics
	and the DFG Priority Program ``Planetary Magnetism'' SPP 1488 for support.
	Computational resources were kindly provided by the PRACE project prj.1602-008
	in the Beskow cluster at the PDC/KTH, Sweden.
	We also used the Hydra cluster of the Max Planck Computing and Data Facility
	(MPCDF, formerly known as RZG) at Garching, Germany.
	We also thank the referees for their valuable suggestions
	that contributed to correct and significantly improve this paper.
\end{acknowledgments}

\appendix
\section{Effects of numerical parameters on
the frequency spectra\label{appendix}}

\begin{figure*}[!ht]
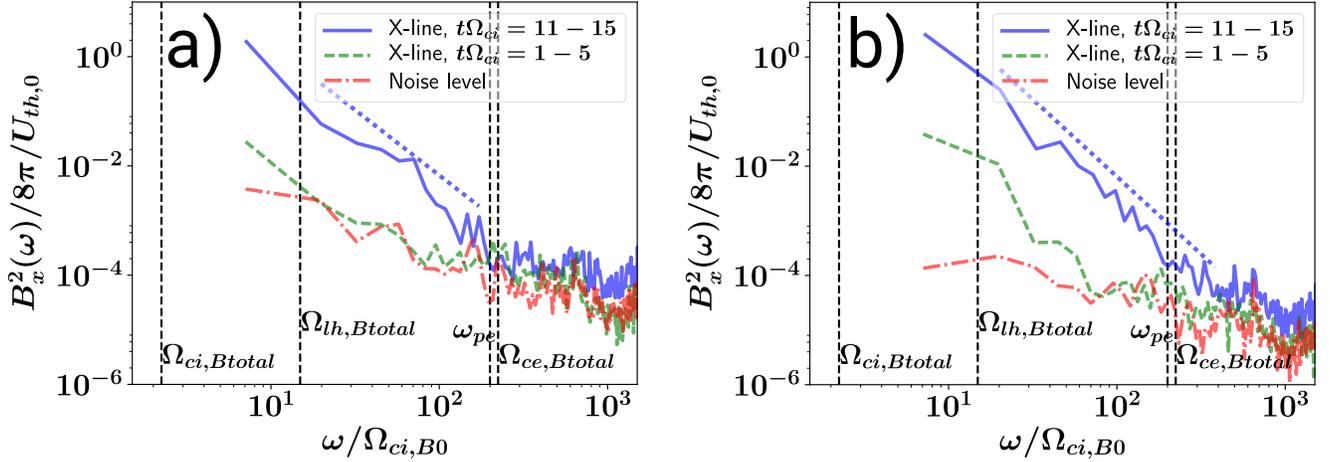
\centering
		\includegraphics[width=0.99\linewidth]{{{fig10_comparison_freq_spectra_ppc}}}
		\caption{Spectrum of perpendicular magnetic field fluctuations $B_x^2$ at the reconnection X-point, for two time intervals.
	a) Run with 25 particles per cell
	b) Run with 100 particles per cell (results used in this paper, c.f. Fig.~\ref{fig:time_spectra_xline}).
		\label{fig:time_spectra_xline_ppc}}
	\end{figure*}

The numerical noise in PIC simulations might have a strong influence
on the turbulence properties of plasmas.
This numerical noise depends on parameters like the shape function
(interpolation scheme to assign the particles' current to the grid),
current smoothing and specially on the number of particles per cell.
We used a second order TSC (triangular shaped cloud) shape function and a binomial current smoothing to reduce the numerical noise.
We also tested that even for four times less particles per cell than
the number used here, the frequency spectral index in Fig.~\ref{fig:time_spectra_xline}
is not modified significantly.
Indeed, Fig.~\ref{fig:time_spectra_xline_ppc} shows a comparison of the frequency spectra
for runs different in only the number of particles per cell.
A higher particle count number
leads to an even clearer spectral slope with only a slightly different value of the spectral index (-2.8 vs -2.7 in the case with less particles) by increasing the turbulent range before it hits the numerical noise floor at high (electron) frequencies.
This numerical noise floor is of course lower when using a higher number of particles.

We also checked that the choice of the time series interval used for the calculation
of the frequency spectra has only a slight effect on the spectral index of magnetic
fluctuations. Indeed, the spectral index calculated for other intervals like
$t\Omega_{ci}=9-14$ (instead of $t\Omega_{ci}=11-15$ used here) is $-2.7$.
Therefore, as long as the interval includes the times close to $t\Omega_{ci}=13.5$,
the spectral index is not very sensitive to the choice of time interval
for the frequency spectra. We finally chose $t\Omega_{ci}=11-15$ in order to
include times close to the maximum value of the reconnection rate.

There is another numerical parameter that can affect the value of the
spectral index of magnetic fluctuations in the frequency domain:
the range used for the linear fit. The lower limit depends on the time series
interval: a larger time interval implies the possibility of choosing an even lower frequency limit,
but this is constrained by the transient nature of magnetic reconnection
in our system, since it would not be meaningful to choose an extended time
interval where reconnection is absent (like at the beginning of our simulation).
The lower limit for the fitting also depends on the amount of binning used
to smooth the data. Binning and averaging over more data points lead to a
smoother spectra but also the lower part of the frequency spectra becomes modified.
The values finally chosen in this paper represent a good compromise between those
opposite effects.
Meanwhile, the upper limit for the range of the linear fit depends on the level
of numerical noise. A higher level of numerical noise, like in the case of using
less numerical particles, implies that the noise level is higher and therefore
the range where a straight line can be fitted in the frequency spectra
is shorter. The same effect happens with reduced or no binning used for smoothing
the input data. For example, we used here as upper limit $\Omega_{ci,B0}=400$,
while without binning (raw data) it is more appropriate to use $\Omega_{ci,B0}=250$.

\end{document}